\documentclass[a4paper,fleqn]{cas-dc}

\usepackage[authoryear]{natbib}
\usepackage{caption}
\usepackage{booktabs} 
\usepackage{siunitx}
\usepackage{hyperref}
\usepackage{fontawesome5} 
\usepackage{xcolor}       

\definecolor{orcidgreen}{HTML}{A6CE39}

\def\tsc#1{\csdef{#1}{\textsc{\lowercase{#1}}\xspace}}
\tsc{WGM}
\tsc{QE}

\begin{document}
\let\WriteBookmarks\relax
\def\floatpagepagefraction{1}
\def\textpagefraction{.001}

\shorttitle{}    

\shortauthors{C.S. Duffey et~al.}  

\title [mode = title]{Detection of Organics in Water Ice by Optical-PhotoThermal Infrared Spectroscopy}  

\tnotemark[1] 

\tnotetext[1]{} 

%

\author[1,2]{Christopher S. Duffey\,\href{https://orcid.org/0009-0000-9901-5269}{\textcolor{orcidgreen}{\faOrcid}}}[type=editor,
                        auid=000,bioid=1,
                        prefix=,
                        role=,
                        orcid=0009-0000-9901-5269]                       
\credit{Conceptualization, Methodology, Investigation, Software, Formal Analysis, Writing - Original draft preparation}

\cormark[1]

\fnmark[1]

\ead{christopher.duffey@ucf.edu}

\ead[url]{}

\credit{}

\affiliation[1]{organization={University of Central Florida},
                addressline={4000 Central Florida Blvd.}, 
                city={Orlando},
                postcode={32816}, 
                state={Florida},
                country={USA}}

\author[1,2]{Julie Brisset}[type=editor,
                        auid=000,bioid=1,
                        prefix=,
                        role=,
                        orcid=0000-0003-2627-335X]  
\credit{Conceptualization, Supervision, Funding acquisition, Writing - Review \& editing}

\author[1,2]{Myles Hoskinson}[type=editor,
                        auid=001,bioid=2,
                        prefix=,
                        role=,
                        orcid=0009-0008-2228-2391]
\credit{Resources, Investigation}                        

\author[2]{Jakob Haynes}[type=editor,
                        auid=002,bioid=3,
                        prefix=,
                        role=,
                        orcid=0009-0000-7491-0814]
\credit{Resources, Investigation}                         
                        
\author[2]{Kathrine Lyakh}[type=editor,
                        auid=003,bioid=4,
                        prefix=,
                        role=,
                        orcid=0009-0004-6985-9187]
\credit{Resources, Investigation}

\fnmark[2]

\ead{}

\ead[url]{}

\affiliation[2]{organization={Florida Space Institute},
                addressline={12354 Research Parkway}, 
                city={Orlando},
                postcode={32816}, 
                state={Florida},
                country={USA}}


\fntext[1]{}


\begin{abstract}
 The detection and characterization of organic molecules within water ice is central to the interpretation of observations of icy moons, comets, and other volatile-rich planetary bodies. However, interactions between organics and the surrounding ice matrix can modify intrinsic water-ice absorption features, complicating the interpretation of infrared spectra. In this study, Optical Photothermal Infrared (O-PTIR) spectroscopy was evaluated as a non-destructive, sub-micrometer resolution technique for the detection and quantification of organics embedded within water ice.

Frozen mixtures containing amino acids and hydroxy acids were analyzed across a wide concentration range using a temperature-controlled sample environment developed for stable ice measurements. L-glycine was detected in water ice down to concentrations of approximately $10^{-6}$ M, with an experimentally determined limit of detection (LOD) of $0.2~\mu$M and a limit of quantification (LOQ) of $3.83~\mu$M. O-PTIR measurements of binary mixtures of L-glycine and lactic acid demonstrated the ability to distinguish and quantify multiple organic components within a single ice matrix.

A broad absorption feature centered near the water bending mode was observed in organic-bearing ice samples but was absent in pure water and deuterium oxide controls. Comparative measurements in H$_2$O and D$_2$O indicate that embedded organics perturb intrinsic ice absorption features, providing new constraints on organic-ice interactions and their influence on infrared spectral interpretation.

These results demonstrate that O-PTIR spectroscopy is capable of detecting, quantifying, and spatially resolving organic compounds within water ice while preserving sample context. The technique offers a promising framework for future laboratory investigations of icy planetary materials and may inform the interpretation of spectroscopic observations of volatile-rich environments.
\end{abstract}


\begin{highlights}
\item{O-PTIR, as a tool for astrobiology, provides sub-micron, non-destructive mapping of organics in water ice.}
\item{High spectral and spatial resolution preserves organic/ice contextual relationship}
\item{Embedded organics perturb intrinsic water ice infrared absorption bands.}
\item{Achieved a 0.2 µM limit of detection for L-glycine in water ice.}
\item{Multiple organic components were quantified within a single ice matrix.}
\end{highlights}



\maketitle

\section{Introduction}

\subsection{Icy Bodies and Astrobiology}

Ocean worlds and icy bodies within our solar system, most notably Jupiter’s moon Europa, Saturn’s moon Enceladus, and various cometary bodies, have emerged as important target locations in the search for life beyond Earth. The Planetary Science and Astrobiology Decadal Survey 2023–2032 explicitly prioritizes the exploration of these potentially organic rich environments, underscoring their potential for significant astrobiological discoveries \citep{NASEM2022}.

To evaluate the habitability of these worlds and unravel the chemical origins of life, the in-situ detection of organic molecules embedded within surface ices is critical \citep{Hand2017}. Analyzing these samples directly on the surface provides pristine, unaltered data, bypassing the sample return challenges, and the potential for terrestrial contamination \citep{Neveu2018}. Furthermore, understanding the formation of organics requires more than just identifying chemical formulas; it requires preservation of the precise spatial relationships between the mineralogy, organic compounds, and surrounding volatiles during analysis. This spatial context may provide additional clues regarding the geochemical processes driving prebiotic chemistry \citep{Cable2012}.

The relevance of these targets is underscored by recent cosmochemical discoveries. Amino acids, the foundational building blocks of proteins, have been consistently identified across diverse celestial bodies. For instance, glycine and alanine represent the most commonly detected amino acids in both carbonaceous meteorites and pristine extraterrestrial samples (\cite{GLAVIN2018205}). The Rosetta mission directly probed comet 67P/Churyumov–Gerasimenko and detected volatile glycine \citep{refId0} and isotopic analysis revealed that glycine detected in the comet Wild 2 samples returned in the Stardust mission were of extraterrestrial origin \citep{Elsila2009}, while the recent OSIRIS-REx sample return mission confirmed the presence of amino acids like glycine and alanine on the carbonaceous asteroid Bennu \citep{doi:10.1073/pnas.2517723123}. Additionally, hydroxy acids such as lactic acid have been detected across multiple carbonaceous chondrites \citep{Pizzarello2010}. Lactic acid is of particular interest because it can facilitate oligomerization reactions between amino acids at low temperatures, offering a viable pathway for complex polymer synthesis in cold planetary environments \citep{Cleaves2008}.

Despite these advances, a fundamental challenge remains in the interpretation of infrared observations of icy planetary surfaces. Absorption features attributed to water ice are commonly used to infer physical properties such as temperature, crystallinity, and grain structure \citep{Boogert2015}. However, these same features are also sensitive to the presence of embedded organic and volatile species \citep{Gerakines1995,Boogert2015}. Interactions between organics and the surrounding ice matrix can modify band positions, broaden spectral features, and alter relative intensities \citep{Hudson2001,Boogert2015}. Resolving this ambiguity requires laboratory measurements that isolate intrinsic absorption processes at the scale of compositionally homogeneous ice domains, where the respective contributions of physical structure and molecular composition can be disentangled through direct spectral response. The increasing sensitivity of modern infrared observatories, including JWST, further emphasizes the need for laboratory constraints to interpret complex and spatially unresolved ice spectra \citep{sturm2024jwst}.

\subsection{Analytical Challenge and Chosen Technique}

Characterization of organic molecules in ices presents multiple analytical challenges. Traditional in-situ techniques often rely on destructive sample preparation (such as pyrolytic volatilization for gas chromatography-mass spectrometry), which destroys spatial context and potentially alter organic structures \citep{Mahaffy2012}. Conversely, non-destructive optical methods like conventional Fourier Transform infrared (FTIR) spectroscopy are fundamentally limited by optical diffraction, restricting spatial resolution to several microns which may be too coarse to map sub-micron organic-mineral interfaces \citep{Centrone2015}.

In this work, Optical Photothermal Infrared (O-PTIR) spectroscopy is applied to address these limitations in the context of planetary ice analogs \citep{COX2025106101,Prater2024}. O-PTIR overcomes the traditional IR diffraction limit by utilizing a sub-micron visible probe laser to detect photothermal expansions induced by a co-aligned, tuneable IR pump laser. This technique offers several distinct advantages for spaceflight instrument suites:
\begin{itemize}
\item 
\textbf{Non-destructive Analysis:}Sub-micron, spectroscopic material mapping can be accomplished using non-destructive laser powers that will ensure the thermal and physical integrity of the samples is maintained during measurement.
\item 
\textbf{Sub-micron Spatial Resolution:}By decoupling the spatial resolution from the IR wavelength, O-PTIR enables the visualization of organic distribution at the sub-micron scale, directly preserving the spatial architecture of organics, volatiles, and minerals.
\item 
\textbf{SWaP Optimization:}The optical architecture of O-PTIR systems lends itself to substantial Size, Weight, and Power (SWaP) reductions, facilitating the miniaturization required to meet stringent payload constraints for future planetary landers and orbital rendezvous missions around small bodies.

\end{itemize}

\subsection{Research Objectives}

The primary objective of this study is to evaluate the viability of O-PTIR for future planetary exploration missions by characterizing analog ice samples in a laboratory environment. Specifically, this work aims to:

\begin{itemize}
    \item Establish the detection limits of O-PTIR for trace astrobiologically relevant organic molecules embedded within a bulk water-ice matrix.

    \item Investigate the spectral characteristics and quantitative capabilities of the technique when analyzing complex, multi-component organic ice matrices.
\end{itemize}

These objectives are addressed through the systematic characterization of laboratory-standard single and binary mixtures of organic species in water ice. In particular, perturbations to the water bending mode, including shifts in peak position and spectral broadening, provide a sensitive probe of coupling between organics and the ice matrix. These effects offer a direct observable of organic-induced modification of intrinsic ice absorption features, enabling improved interpretation of infrared spectra of icy planetary environments.

\section{Methods}

\subsection{O-PTIR Instrumentation and Environmental Control}
The core of the analysis utilizes data collected with an Optical Photothermal Infrared (O-PTIR) spectroscopy system, which is capable of collecting mid-wave infrared (IR) absorption spectra across the $5.4 - 10.4~\mu m$ range. A key feature of this instrument is its ability to perform measurements with sub-micron spatial resolution. The O-PTIR method employs low laser powers (less than $20~\text{mW}$ for Mid-wave IR and less than $4~\text{mW}$ for Visible) to enable non-destructive measurement, which is crucial for collecting data from sensitive ice surfaces with minimal to no alteration.
To facilitate long-term stability for ice sample analysis, two major modifications/refinements were introduced to the standard instrument:

\begin{itemize}
\item{\textbf{Peltier Cooling System}: A custom-designed, 3-stage Peltier cooling system was developed and integrated beneath the sample holder. This system is engineered to rapidly freeze organic/ice solutions directly inside the instrument and maintain a constant sub-zero temperature over collection periods spanning minutes to hours. The design ensures that it does not interfere with the normal optical operation of the O-PTIR instrument and prevents the transmission of liquid pump vibrations into the sample.}
\item{\textbf{Atmospheric Frost Mitigation}: Ambient water vapor quickly forms a thick layer of frost on cold ice samples, which severely interferes with the collection of accurate spectra from the organic-ice surface. To counteract this, the instrument chamber employs an active instrument purge system. By maintaining the air within the sample chamber at a constant relative humidity (RH) of $8\%$ or less, frost formation is mitigated such that it does not impact the experimental data.}
\end{itemize}

\subsection{Sample Preparation and Storage Protocol}
\subsubsection{Single-Component Standards Preparation}
The amino acid L-glycine (in dry crystalline form) was used to create a series of single-component solution standards, prepared by serial dilution in pure de-ionized (DI) water across concentrations from $0.1\text{M}$ down to $10^{-6}\text{M}$. Strict cleanliness protocols were followed to prevent contamination of the samples during all phases of the experiment, this included during sample preparation, introduction of the samples into the instrument and long term refrigerated storage of the samples.

Additionally, the following single-component comparison standards were prepared:
\begin{itemize}
\item{L-glycine in dry crystalline powder form.}
\item{L-alanine in dry crystalline powder form.}
\item{Lactic acid in dry crystalline powder form.}
\item{A $0.1\text{M}$ solution of lactic acid in DI water.}
\item{A $0.1\text{M}$ L-glycine solution in deuterium oxide ($\text{D}_{2}\text{O}$).}
\item{A $0.023\text{M}$ solution of L-alanine in DI water.}
\item{Control sample of pure de-ionized water.}
\item{Control sample of pure deuterium oxide ($\text{D}_{2}\text{O}$).}
\end{itemize}
These single component standards were used to validate the O-PTIR instruments absorption spectra accuracy against existing standards and databases collected using FTIR or other older methods as well as providing background control spectra of pure de-ionized water and deuterium oxide.

\subsubsection{Binary Mixture Standards Preparation}
For quantitative analysis, binary mixtures of L-glycine (Organic A) and lactic acid (Organic B) were prepared in de-ionized water. The parts per million (ppm) amount of each component was adjusted to control the mixture ratios, holding the total molarity at a constant of 0.1M for the solution, thus the L-glycine-to-lactic acid ratio (A:B) was systematically varied across a calibration series of five standards:

\begin{itemize}
\item{$90\% \text{   L-glycine} : 10\% \text{ Lactic Acid}$}
\item{$75\% \text{   L-glycine} : 25\% \text{ Lactic Acid}$}
\item{$50\% \text{   L-glycine} : 50\% \text{ Lactic Acid}$}
\item{$25\% \text{   L-glycine} : 75\% \text{ Lactic Acid}$}
\item{$10\% \text{   L-glycine} : 90\% \text{ Lactic Acid}$}
\end{itemize}

\subsubsection{Sample Storage Protocol}
After the liquid samples are created they are immediately refrigerated to \qty{2}{\celsius} for storage, this helps to inhibit microbial growth and also shortens the time to freeze significantly when placed in the cooled sample holder of the O-PTIR instrument.

\subsection{Data Acquisition and Analysis}
\subsubsection{Data Acquisition}
The data acquisition process begins with the acclimatization of the O-PTIR instrument to bring the internal relative humidity down and stabilize the sample cooling system.  The active sample cooling utilizes multiple Peltier coolers, an external liquid cooling loop, radiator, and an ice bath to bring the sample holder to a stable \qty{-15}{\celsius}. Figure~\ref{fig:Figure_1} shows the custom cooler through the open access door on the O-PTIR instrument.

\begin{center}
    \includegraphics[width=0.48\textwidth, alt={O-PTIR instrument door open shhowing the modified sample stage with additional cooling block, flexible liquid cooling hoses visible}]{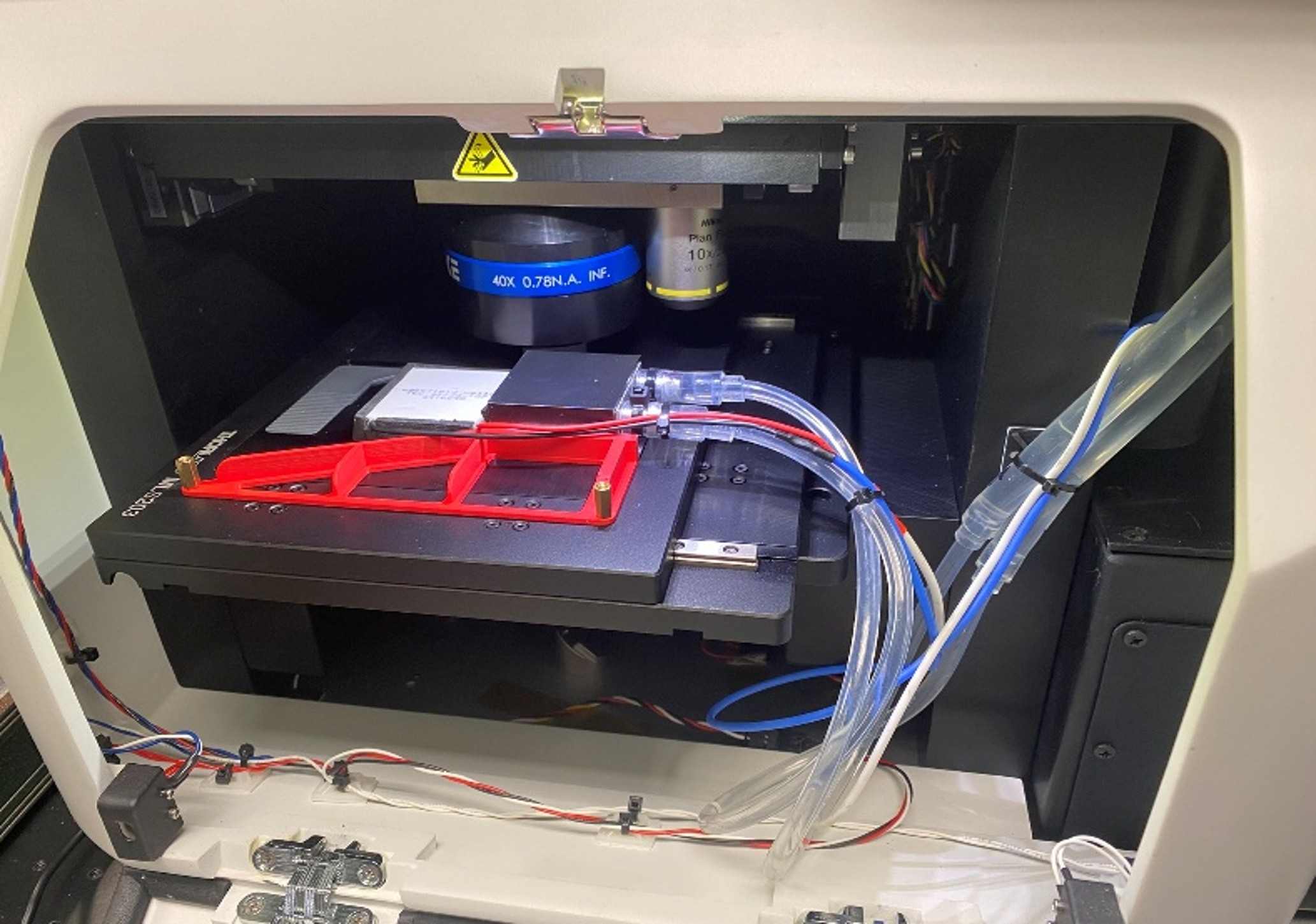}
    \captionof{figure}{Sample Cooler Stage in the O-PTIR Instrument}
    \label{fig:Figure_1}
\end{center}

Once the Instrument internal humidity has achieved less than $8\%$ RH the instrument background calibration is performed using a built in low emissivity target. This procedure will correct the data for the remaining small amount of atmospheric water as well as any amplitude/wavelength variations in the quantum cascade laser (QCL). Once the instrument is ready for collection, a the sample of the liquid solution to be analyzed is removed from refrigeration and a small drop is placed in the instrument sample holder.  When the sample reaches a stable frozen state, multiple hyperspectral maps were collected ($120 \times 120~\mu m$ with $7 \times 7$ points for 49 spectra). The hyperspectral maps were used to select regions of interest for analysis and verify the spatial distribution of the organics. Figure~\ref{fig:Figure_2} shows the sample frozen in-situ

\begin{center}
    \includegraphics[width=0.48\textwidth]{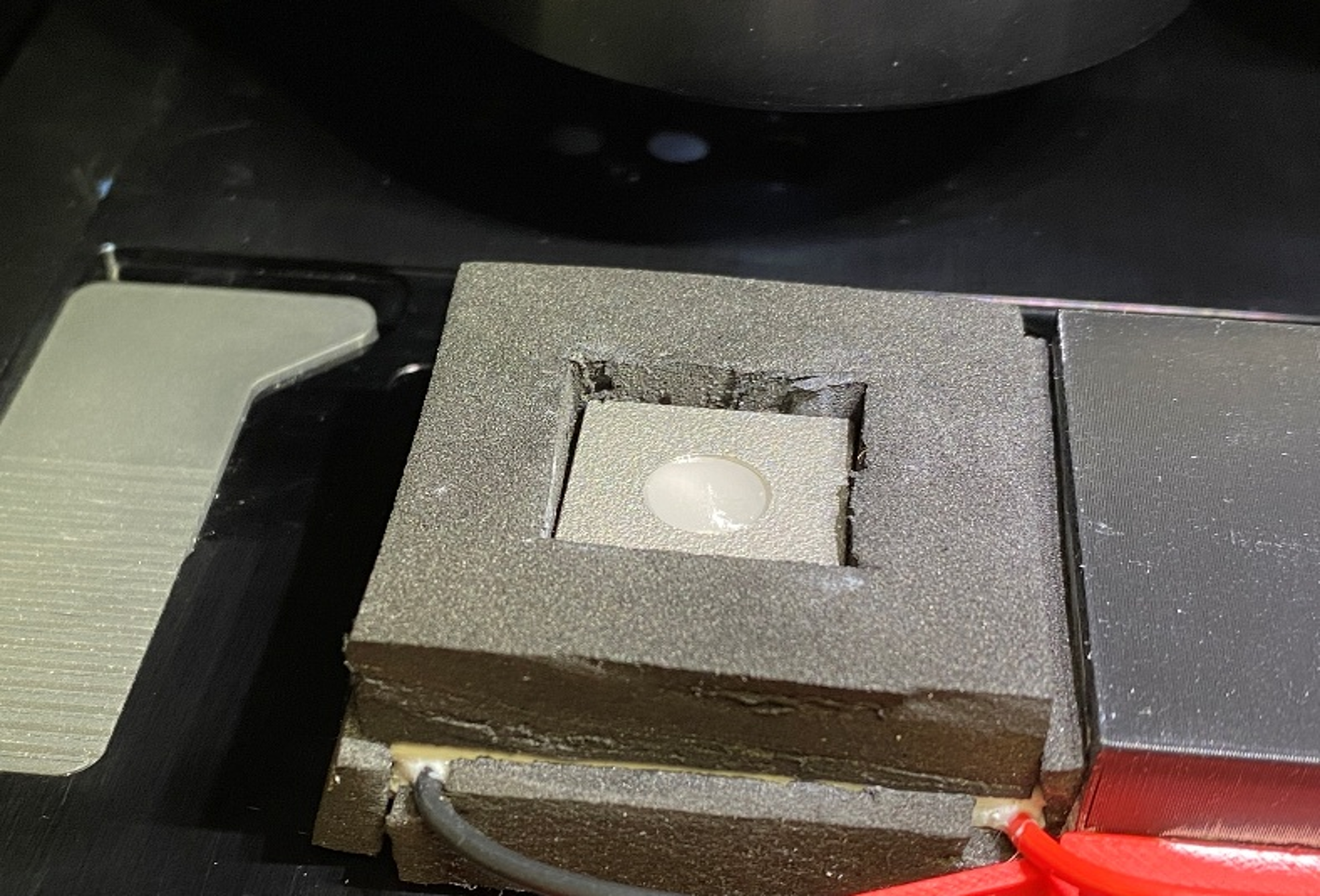}
    \captionof{figure}{Frozen Sample Ready for Spectral Mapping in the O-PTIR Instrument}
    \label{fig:Figure_2}
\end{center}

\subsubsection{Data Analysis Approach}
Two separate analysis were performed on the collected hyperspectral data sets. For the L-glycine and de-ionized water single component standards, spectral data analysis and processing were executed using custom Python scripts to evaluate the spectral match between the experimental samples and a pure crystalline L-glycine reference. 
To quantify the presence of L-glycine within the de-ionized ice matrix, a spectral vector approach was implemented using cosine similarity. The raw intensity values across the target wavenumber range are treated as multi-dimensional vectors. By calculating the scalar product of the normalized sample and reference vectors (the organic standard) we can assess the similarity of the samples. Thus the cosine similarity metric effectively evaluates the structural profile of the resonant peaks independent of absolute scale variations. 
The detection limit was established by tracking the degradation of this similarity score across the dilution series, identifying the minimum concentration where the characteristic profile of the glycine bands remained statistically distinguishable from the ambient background of the water matrix and producing a signal to noise ratio (SNR) vs molarity plot.  Additionally the data was processed to find the Limit of Detection (LOD) and the Limit of Quantification (LOQ), for this approach we applied the established guidelines of $3.3\sigma$ to determine the LOD and $10\sigma$ to determine the LOQ. The method that was implemented to evaluate SNR, LOD, and LOQ in our Python analysis code is described as follows:

\textbf{1. Local Noise and Peak Signal Characterization} Rather than assuming a uniform baseline variance across the full spectrum, a localized noise floor calculation is employed to accurately capture the stochastic fluctuations of the ice mixture. The background standard deviation, $\sigma_{\text{noise}}$, is defined as the root-mean-square (RMS) noise calculated within a blank-subtracted spectral window devoid of spectral features:$$\sigma_{\text{noise}} = \sqrt{\frac{1}{N-1} \sum_{i=1}^{N} \left( I_{\text{sub}}( \nu_i ) - \bar{I}_{\text{sub}} \right)^2}$$where $I_{\text{sub}}(\nu_i)$ represents the blank-subtracted normalized intensity at wavenumber $\nu_i$ within the the reference window ($1750 \text{ cm}^{-1} \le \nu \le 1800 \text{ cm}^{-1}$), and $\bar{I}_{\text{sub}}$ is the mean intensity across those $N$ channels. The net signal level for a given concentration, $S(C)$, is determined by the maximum baseline-corrected intensity of the primary diagnostic fingerprint feature (in this case the large L-glycine peak near $\sim 1610~cm^{-1}$) :$$S(C) = \max \left[ I_{\text{sub}}(\nu) \right] \quad \text{for } \nu \in [\nu_{\text{low}}, \nu_{\text{high}}]$$The experimental, localized signal-to-noise ratio ($\text{SNR}_{\text{local}}$) as a function of molar concentration $C$ is thus formulated as:$$\text{SNR}_{\text{local}}(C) = \frac{S(C)}{\sigma_{\text{noise}}}$$

\textbf{2. Log-Space Interpolation Framework} Because the spectroscopic response spans multiple orders of magnitude across the sample molarity regime, a standard ordinary least-squares linear regression fails to capture the localized nonlinearities of the organic/ice mixture hardware response. To evaluate sensitivity thresholds without forcing a global linear fit over a wide dynamic range, the detection limits are calculated using localized log-space piecewise interpolation. The independent variable is transformed to a logarithmic coordinate frame to linearize concentration-dependent scaling factors:$$x = \log_{10}(C)$$A localized mapping function, $f$, is constructed across adjacent experimental coordinates such that:$$x = f(\text{SNR}_{\text{local}})$$For any target coordinate falling between two adjacent experimental points $(x_1, \text{SNR}_1)$ and $(x_2, \text{SNR}_2)$, the localized sensitivity slope ($\alpha_{\text{local}}$) is defined as:$$\alpha_{\text{local}} = \frac{\text{SNR}_2 - \text{SNR}_1}{x_2 - x_1} = \frac{\Delta \text{SNR}_{\text{local}}}{\Delta \log_{10}(C)}$$The continuous log-concentration coordinate for a specific target threshold $T$ is mathematically evaluated via:$$x_T = x_1 + \frac{T - \text{SNR}_1}{\alpha_{\text{local}}}$$

\textbf{3. Threshold Evaluation (LOD and LOQ)} Consistent with the International Union of Pure and Applied Chemistry (IUPAC) performance metrics (\cite{ich2023q2r2}), the standardized statistical criteria for reliable detection and quantification are defined at multiples of $3.3\sigma$ and $10\sigma$, respectively. The empirical LOD is established at the boundary where the net signal is exactly $3.3$ times greater than the background noise standard deviation ($\text{SNR}_{\text{local}} = 3.3$). The log-space coordinate is calculated by evaluating the localized interpolation matrix at the target threshold:$$x_{\text{LOD}} = f(3.3)$$ Converting out of log-space yields the physical concentration threshold:$$\text{LOD}_{\text{molar}} = 10^{x_{\text{LOD}}} = 10^{f(3.3)}$$ The empirical LOQ marks the strict baseline where the signal is sufficiently isolated from noise variance to permit accurate quantification ($\text{SNR}_{\text{local}} = 10.0$). The threshold is evaluated via:$$x_{\text{LOQ}} = f(10.0)$$$$\text{LOQ}_{\text{molar}} = 10^{x_{\text{LOQ}}} = 10^{f(10.0)}$$ This localized interpolation approach ensures that both LOD and LOQ analytical thresholds are anchored to the actual instrument performance.

Quantitative analysis of the binary mixtures was performed by comparing the 
integrated peak areas of the characteristic spectral features for L-glycine 
and lactic acid within the fingerprint region (1800--900~cm$^{-1}$). 
To evaluate differences from ideal mixing behavior, as described by Beer's Law, a simple model was created from the reference spectrum for each mixture concentration via a linear superposition of the pure crystalline component spectra, weighted by their respective parts-per-million (ppm) fractions. The coefficient of determination ($R^2$) was calculated between the experimental and synthetic spectra to quantify the geometric similarity and identify regions of model divergence. 

To evaluate the linearity of the spectroscopic response, an empirical calibration curve was constructed by plotting the ratio of the true mixture concentrations against the ratio of the integrated peak areas recovered by the O-PTIR spectroscopic analysis, ($A_{\text{lactic}}/A_{\text{glycine}}$) for the characteristic bands at 1760--1680~cm$^{-1}$ and 1150--1080~cm$^{-1}$, respectively. Localized physical perturbations within the crystalline ice matrix were evaluated by calculating a spectral Broadening Factor, defined as the ratio of the Full-Width at Half-Maximum (FWHM) of the characteristic lactic acid feature in the mixture to that of the pure reference sample ($FWHM_{\text{mixture}} / FWHM_{\text{pure}}$). This approach allows for the decoupling of simple concentration scaling from localized intermolecular interactions, such as hydrogen bonding and potentially organic segregation at ice grain boundaries.

\section{Results and Discussion}

\subsection{O-PTIR Spectroscopic Signatures of Single Organics in Ice}

\subsubsection{Spectral Characterization of L-glycine, L-alanine and Lactic Acid Standards}
To validate the efficacy of the O-PTIR approach to collecting a Mid-wave IR absorption spectra from $5.4 - 10.4~\mu m$, we compared the spectral features of crystalline L-glycine, L-alanine, and Lactic Acid against standard databases (Wiley and NIST sources).  The samples of crystalline organics, as expected, exhibit a random spatial orientation in the instrument that is characterized by variations in the peak amplitudes. This effect is mitigated by collecting large numbers of spatially diverse hyperspectral points and computing an average spectral shape. Figure~\ref{fig:Figure_3} shows the variation in amplitudes for multiple L-alanine spectra and their average value. Figure~\ref{fig:Figure_4} shows a comparison of crystalline L-alanine versus a spectral standard for an FTIR absorption spectrum of L-alanine. The agreement between the FTIR and O-PTIR measurements are excellent and demonstrates the applicability of the O-PTIR method to probe for amino acids and other organic molecules.

\begin{center}
    \includegraphics[width=0.48\textwidth]{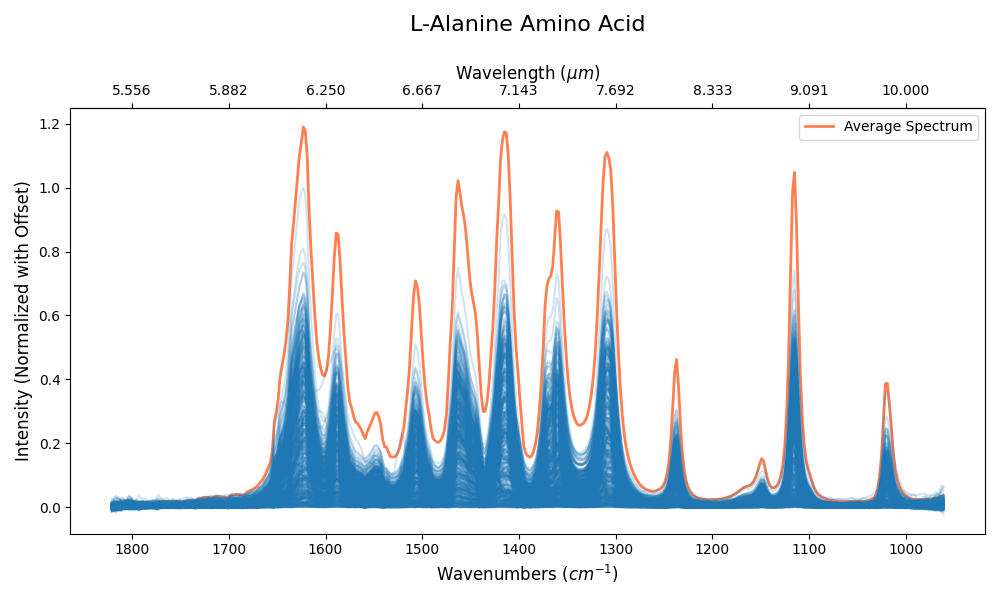}
    \captionof{figure}{Crystalline L-alanine O-PTIR Reference Spectrum}
    \label{fig:Figure_3}
\end{center}

\begin{center}
    \includegraphics[width=0.45\textwidth]{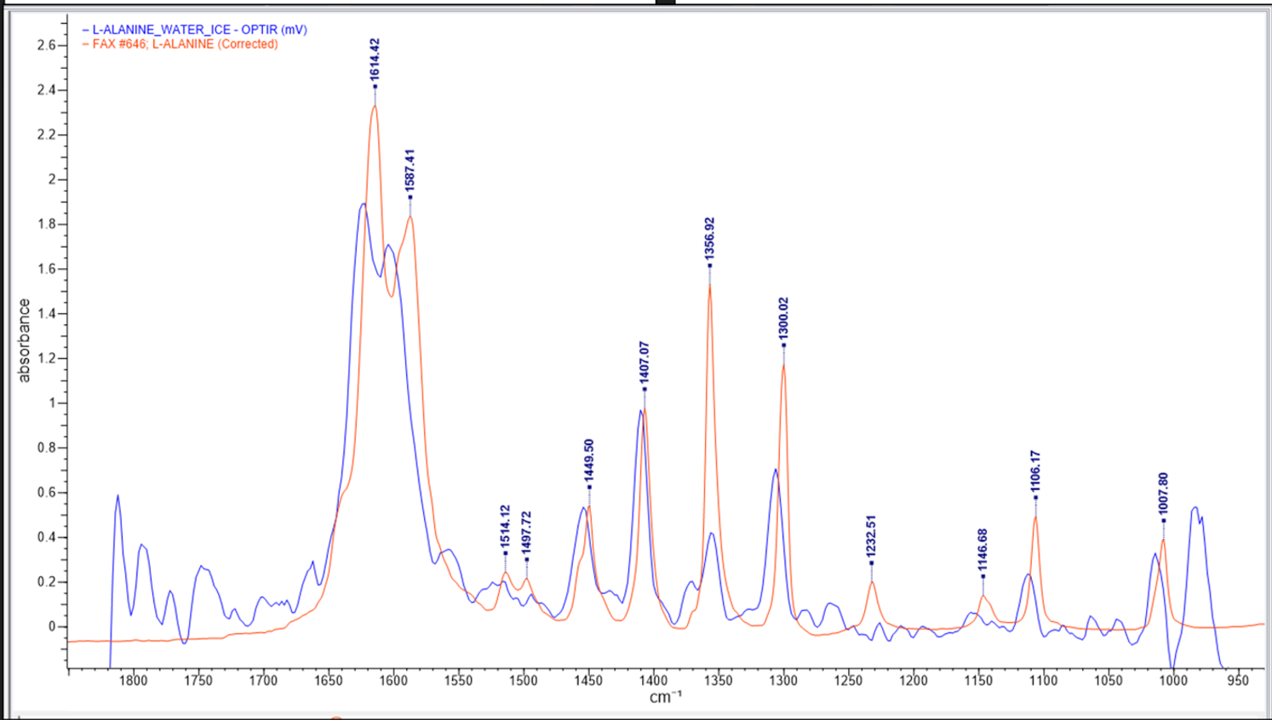}
    \captionof{figure}{Crystalline L-alanine O-PTIR vs Wiley Standard}
    \label{fig:Figure_4}
\end{center}

\subsubsection{Organic Molecules in Water Ice Spectral Data}

To interpret the role of organic–ice interactions in infrared spectroscopy, we first establish the intrinsic spectral response of pure water ice under O-PTIR conditions. This provides a reference baseline that allows separation of absorption features arising from the water matrix itself and those induced by embedded organic species. 

As outlined in Section 1, a central challenge in planetary spectroscopy is the coupling between physical and compositional effects within observed absorption features. By comparing pure ice spectra with organic-bearing mixtures, it becomes possible to isolate perturbations associated with molecular interactions and assess their contribution to the observed spectral response.

For this, we looked at a pure frozen sample of de-ionized water. This spectrum was intended to provide data for background  subtraction from the organic mixes, however we discovered that this pure frozen sample had no significant O-PTIR response in the $5.4 - 10.4~\mu m$ range. Figure~\ref{fig:Figure_5} below shows this flat response with the blue individual spectra and the average in orange. We note that there are tails at the beginning and end of the wavenumber range likely related to the instrument response and there is a slight, broad "hump" seen in a few of the individual spectrum which may be a response to the H-O-H water/ice bending moment at $\sim 1650~cm^{-1}$ (\cite{mate2014glycine}). This featureless behavior indicates that observed absorption features in mixtures primarily arise from organic components and their interaction with the ice matrix.

\begin{center}
    \includegraphics[width=0.48\textwidth]{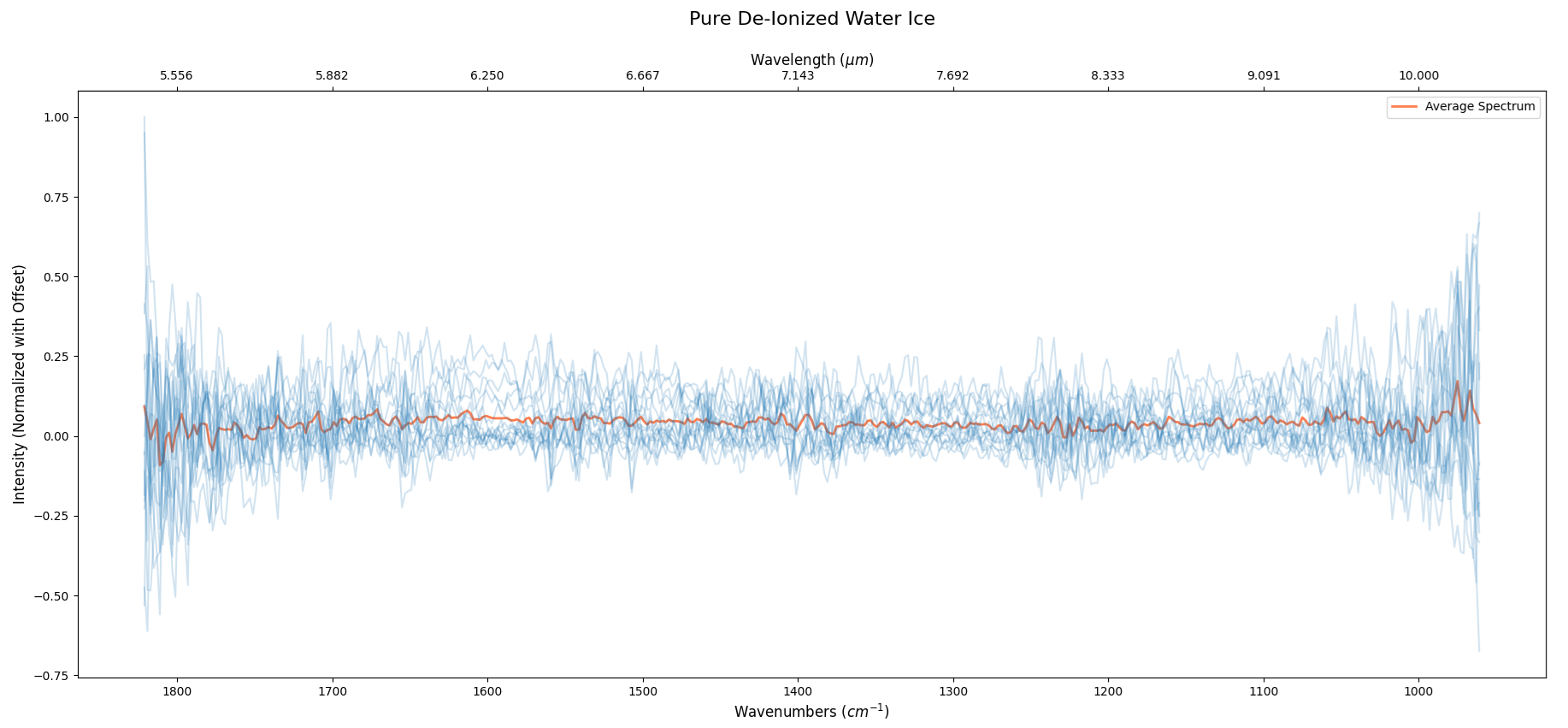}
    \captionof{figure}{Spectrum for Pure Frozen De-ionized Water }
    \label{fig:Figure_5}
\end{center}

Looking at the 0.1M L-glycine water ice mixture in Figure~\ref{fig:Figure_6} we see all the collected individual spectrum in the lower part of the plot (cyan color) we note some amplitude variability similar to what is seen in the crystalline samples, and likely indicative of spatial orientation effects of the molecules to the stimulating laser.  In the upper plot we see the average spectral response of the solution (blue) compared to the crystalline L-glycine (orange). The most striking feature is the dominating broad response around $\sim 1650~cm^{-1}$, other peaks in the response are visible, although reduced in amplitude and in some case broadened as well. The large broad peak is slightly shifted from the L-glycine peak that is at $\sim 1610~cm^{-1}$. This shift is discussed in more detail in section 3.1.3.

\begin{center}
    \includegraphics[width=0.46\textwidth]{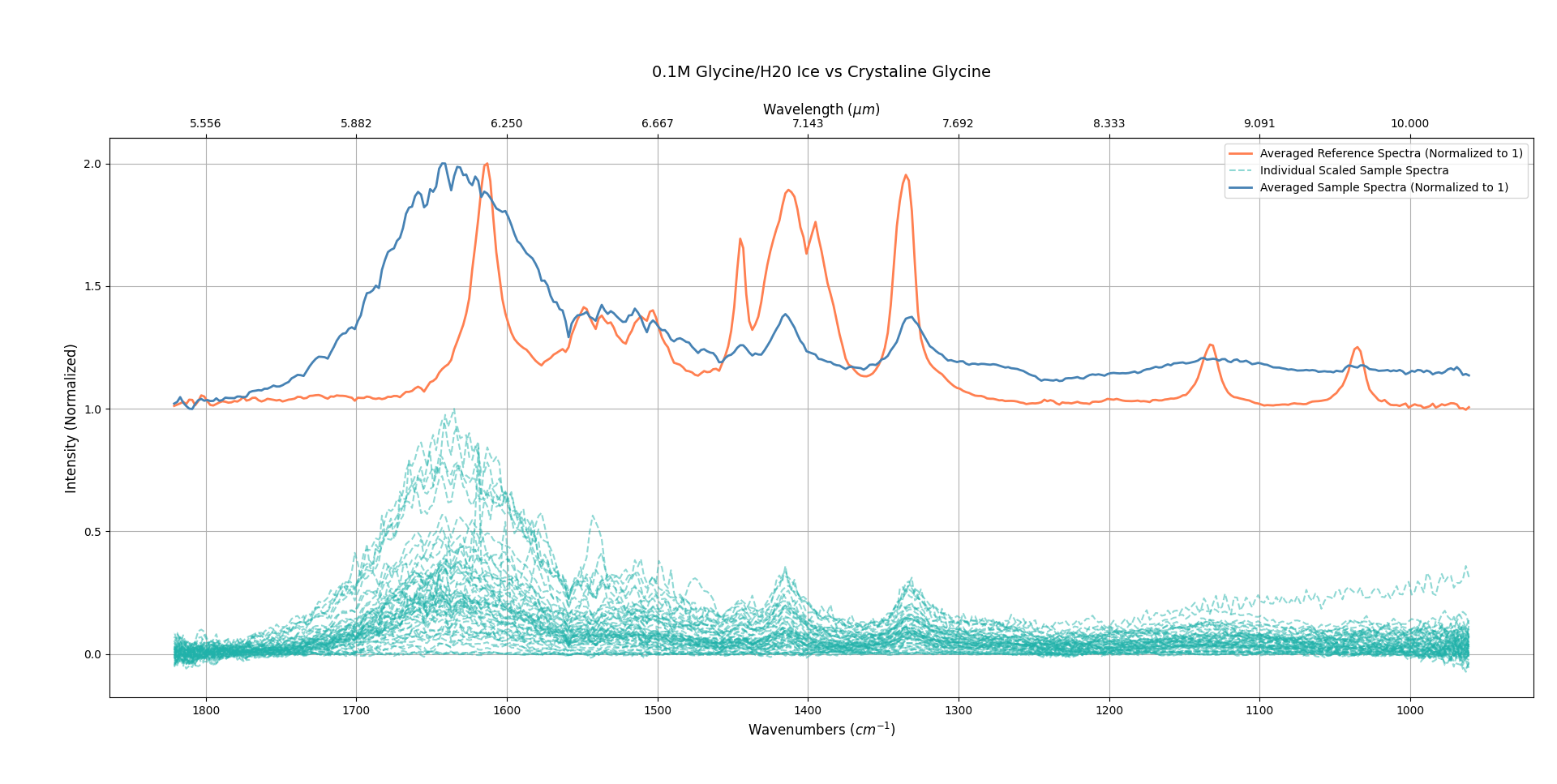}
    \captionof{figure}{Spectrum for 0.1M L-glycine Water Ice }
    \label{fig:Figure_6}
\end{center}

Examining the 0.1M lactic acid and ice mixture in Figure~\ref{fig:Figure_7} we note similar data, although we can see that the broad peak we observed in the L-glycine data has a second peak aligned with the lactic acid peak at $\sim 1745~cm^{-1}$.  The lactic acid also has a very strong peak visible around $\sim 1120~cm^{-1}$ which is a vibrational mode from its methyl group, $CH_{\mathrm{3}}$.

\begin{center}
    \includegraphics[width=0.46\textwidth]{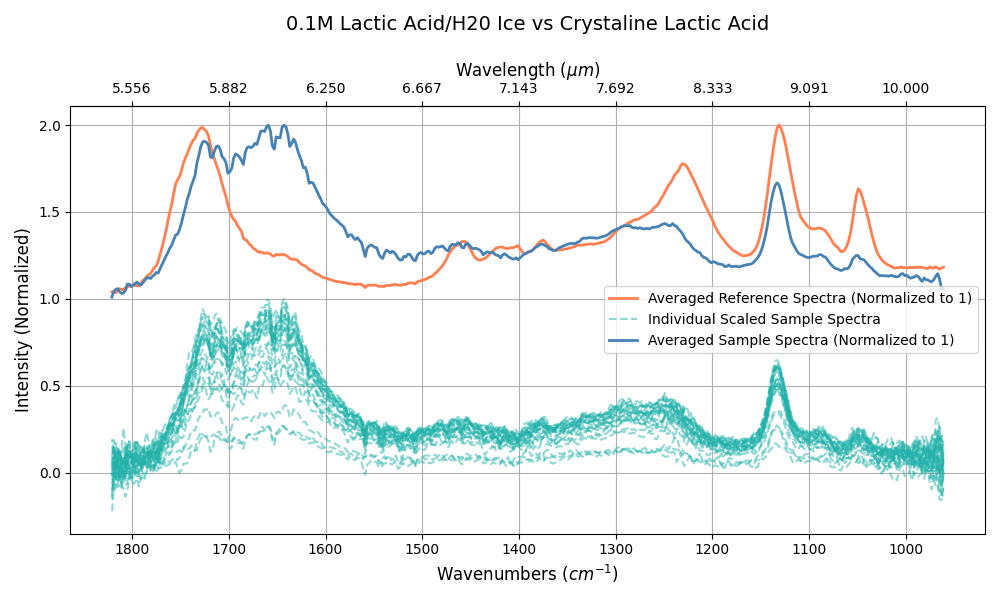}
    \captionof{figure}{Spectrum for 0.1M Lactic Acid Water Ice }
    \label{fig:Figure_7}
\end{center}

\subsubsection{Effects of Deuteration on Organic Peak Wavenumber Shift ($\text{H}_{2}\text{O}$ vs. $\text{D}_{2}\text{O}$)}
Initial observations of the O-PTIR data for frozen L-glycine and de-ionized water solutions showed a consistent shift in the expected large L-glycine absorption peak that should be around $\sim 1610~cm^{-1}$ and shifting to $\sim 1650~cm^{-1}$. This shift happens to be close to the H-O-H bending moment for liquid water/ice $\text{H}_{2}\text{O}$ that occurs at $\sim 1645~cm^{-1}$ suggesting a possible interaction between the organics and the water. To explore this possibility we created a 0.15M test solution with $\text{D}_{2}\text{O}$ instead of $\text{H}_{2}\text{O}$. This modified sample was then imaged in the O-PTIR system.

\begin{center}
    \includegraphics[width=0.48\textwidth]{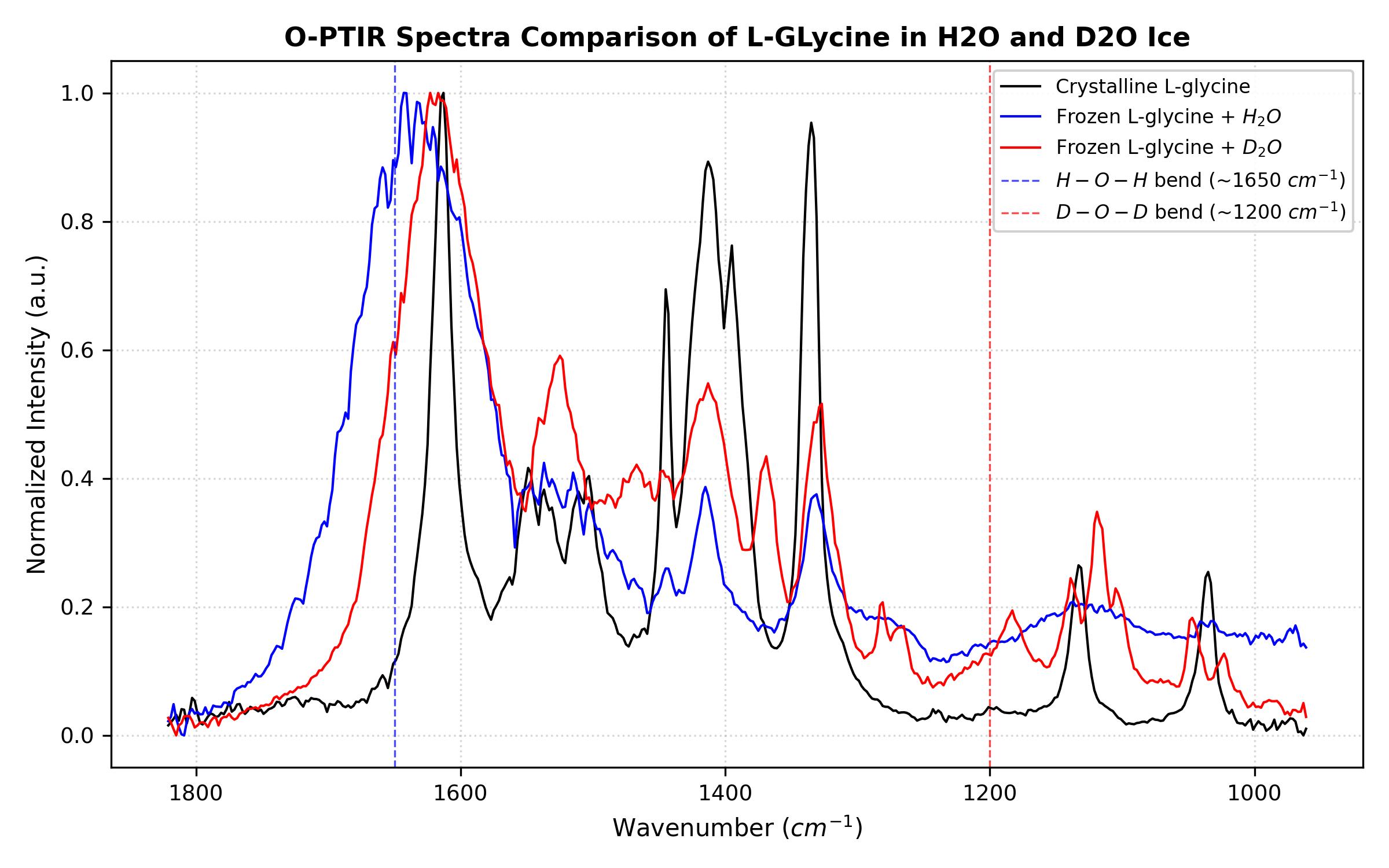}
    \captionof{figure}{Effect of D2O on L-glycine Ice Spectra Peaks}
    \label{fig:Figure_8}
\end{center}

Figure~\ref{fig:Figure_8} Provides evidence that there is an interaction between the water bending moment and the L-glycine, notably causing a shift in the the L-glycine peak that is near the H-O-H bending moment in the de-ionized solution sample.  The L-glycine peak in the $\text{D}_{2}\text{O}$ is now aligned with the crystalline L-glycine peak and also less broadened.  In the $\text{D}_{2}\text{O}$ and L-glycine ice we also see a new spectral peak that is near the D-O-D bending moment - although shifted this may also be an indication of interactions between the heavy water and the L-glycine.  As a control, we collected spectral data of pure frozen $\text{D}_{2}\text{O}$ and was found to be consistent with the data presented for the pure de-ionized water in Figure~\ref{fig:Figure_5}

\subsection{Quantitative Detection Limits in Ice}
\subsubsection{Determination of L-glycine Detection Limit}
Figure~\ref{fig:Figure_9} shows plots of the spectral average for different concentrations ($0.1\text{M}$ down to $10^{-6}\text{M}$) of L-glycine compared with the L-glycine crystalline standard, the background baseline has been subtracted. The errors are shown as the light colored deviations around the averages. Note that as expected, the errors grow larger as the concentration drops. The data is normalized for each concentration and so we see the instrument "tails" discussed in 3.1.2 becoming visible as the SNR drops at the lower concentration values.

\begin{center}
    \includegraphics[width=0.48\textwidth]{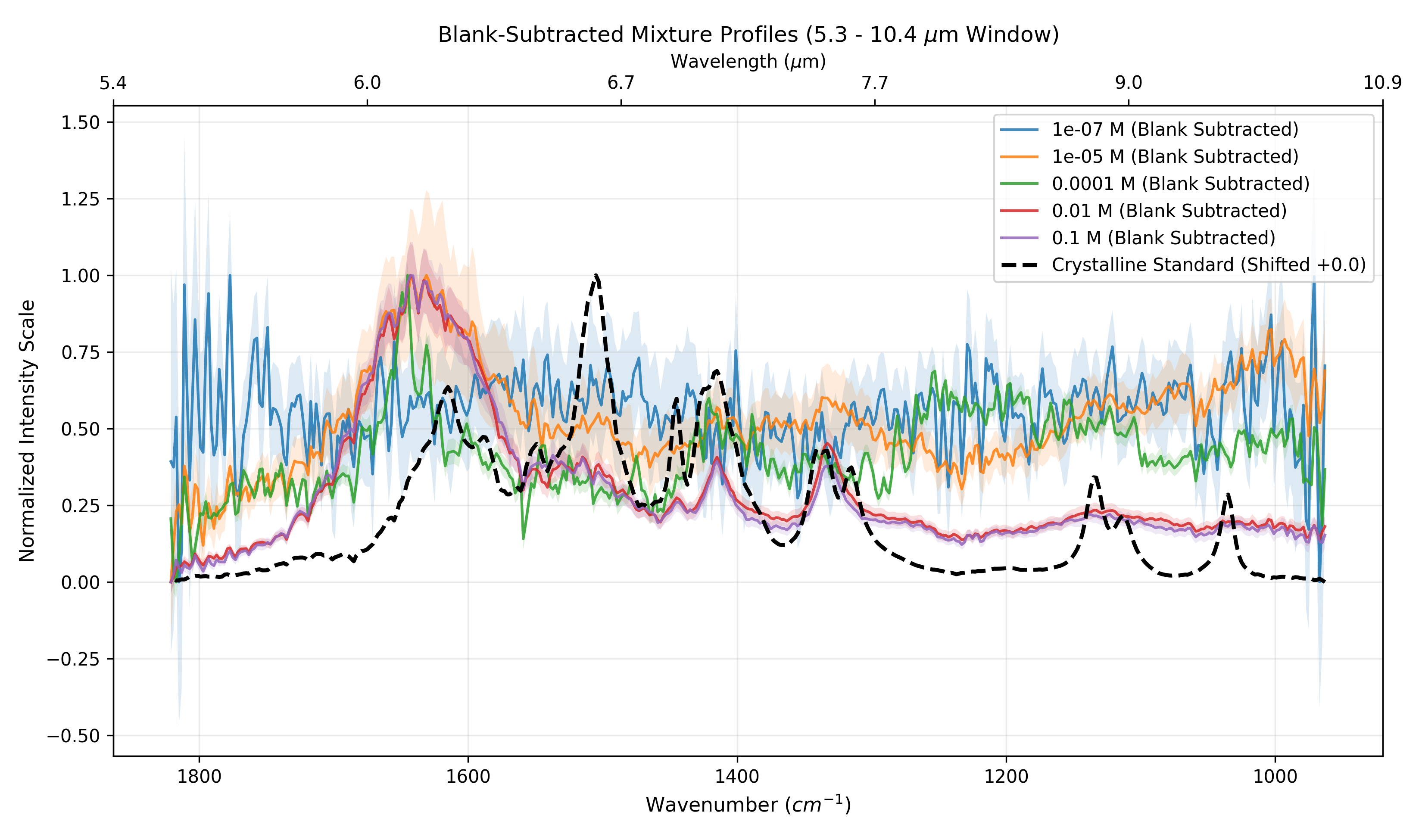}
    \captionof{figure}{L-glycine/Water Ice Mixture Profiles}
    \label{fig:Figure_9}
\end{center}

Figure~\ref{fig:Figure_10} is a plot of the cosine similarity metric of the various mixtures, as compared to the crystalline L-glycine standard, on the left axis.  A similarity value of 1 would be perfect similarity to the standard and a value of 0 would be no similarity.  The similarity metric ranges from a high value of 0.79 to a low of 0.67 and we observe that the error increases from ~0.03 to ~0.20 as the concentration drops below 0.001M. On the right axis we have the computed signal to noise ratio (SNR) vs concentration.  

\begin{center}
    \includegraphics[width=0.48\textwidth]{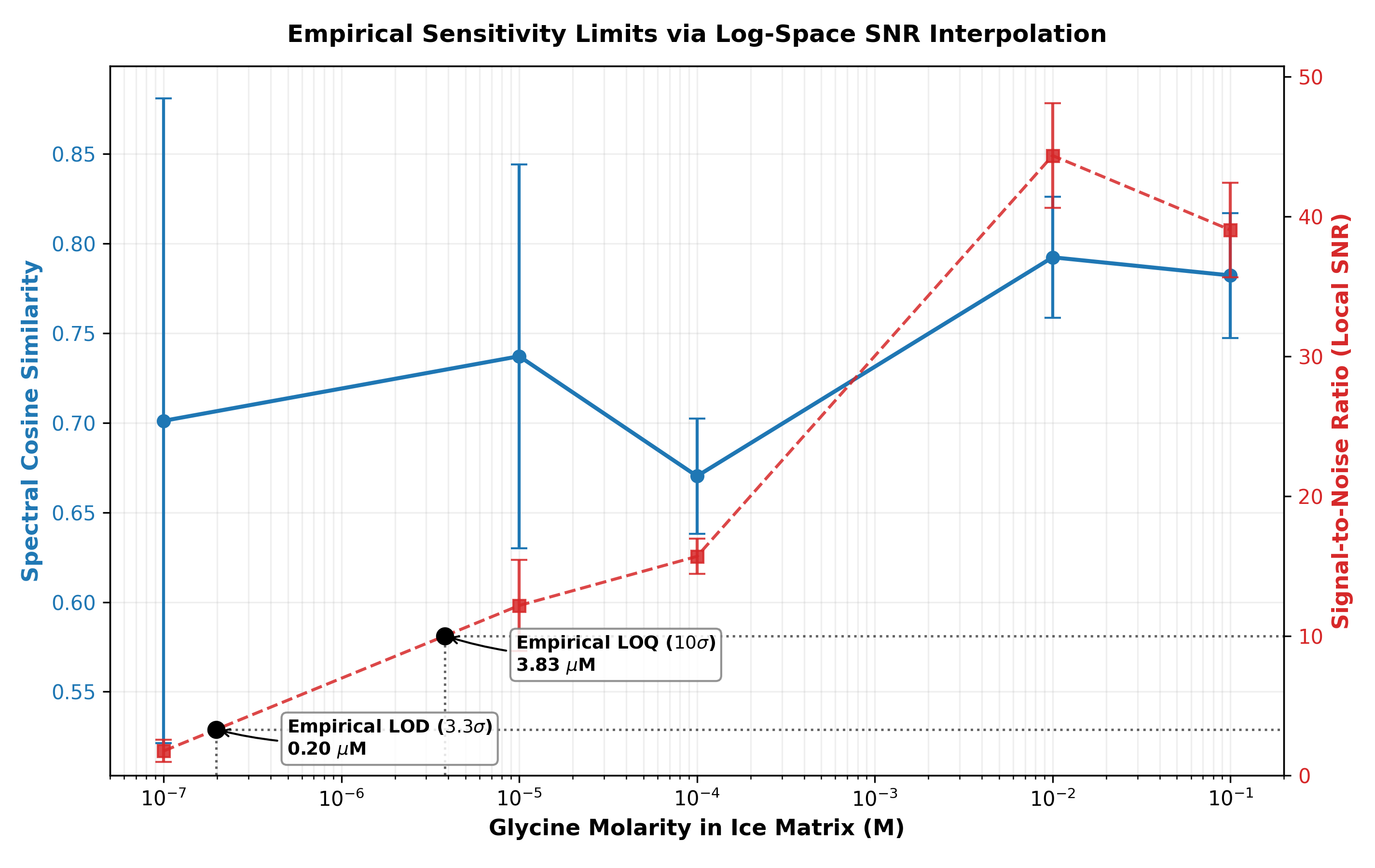}
    \captionof{figure}{L-glycine/Water Ice, SNR and Similarity Metric}
    \label{fig:Figure_10}
\end{center}

Using the methods described in the \textbf{Data Analysis section 2.3.2}, we determine that the limit of detection (LOD) and limit of quantification (LOQ) for L-glycine in water ice measured using O-PTIR is $0.2 \mu M$ (0.0150 ppm) and $3.83 \mu M$ (0.2879 ppm) respectively and represented as the black points on the SNR curve in Figure~\ref{fig:Figure_10}.

\subsubsection{Organic-Ice Phase As Observed Via O-PTIR Spectroscopy}
From the data we see that O-PTIR spectroscopy is able to detect and quantify a mixture of frozen de-ionized water and L-glycine. However we see in all the samples a very broadened spectroscopic feature at/near the H-O-H bending  moment of the pure water \citep{mate2014glycine} which is spectrally collocated with a significant L-glycine feature. We noted that the O-PTIR MW IR absorption spectrum for the control sample of pure frozen water, or deuterium oxide, was flat and featureless. This implies that the feature could be enhanced by the interaction between the organic molecules and the frozen water. 

In \cite{doi:10.26434/chemrxiv.12640760.v1} we see a similar feature at $\sim 1650~cm^{-1}$ for mixtures of pure water and deuterium oxide in liquid form using FTIR. Since this would be analogous to our pure standard, the observed difference between the two methods may be an outcome of the O-PTIR system architecture, the very low laser power levels used, or relate to the way O-PTIR probes the surface of the icy mixtures, i.e. an absorbance spectrum from a reflected probe beam (\cite{Prater2024}). 

It should be noted we also see the broad $\sim 1650~cm^{-1}$ feature with lactic acid that has a large spectral peak more than 100 wavenumbers away from the H-O-H bending moment. The results support some form of vibrational coupling between the water molecules and the organics which are easily stimulated via the O-PTIR low power infrared source and readily detected by the visible probe laser, observationally, as the  concentration of the organics in the ice drops, the strength of this coupling also appears to diminish.

\subsection{Analysis of Binary Organic Mixtures}
\subsubsection{Spectral Resolution of L-glycine/Lactic Acid Mixtures}
In Figure~\ref{fig:Figure_11} we present example O-PTIR spectra for a mixture of 901 ppm Lactic Acid and 6756 ppm of L-glycine. The raw composite hyperspectral data is in the lower panel, showing the spatial variability of each point measured. The upper panel includes the average of the mixture spectra (blue line) and the raw standard spectra for crystalline Lactic Acid (green line) and L-glycine (orange line), all three of these spectra are normalized for easy visual comparison. This data represents the highest L-glycine to lactic acid mixture ratio and the data is easily interpretable as showing that.

\begin{center}
    \includegraphics[width=0.48\textwidth]{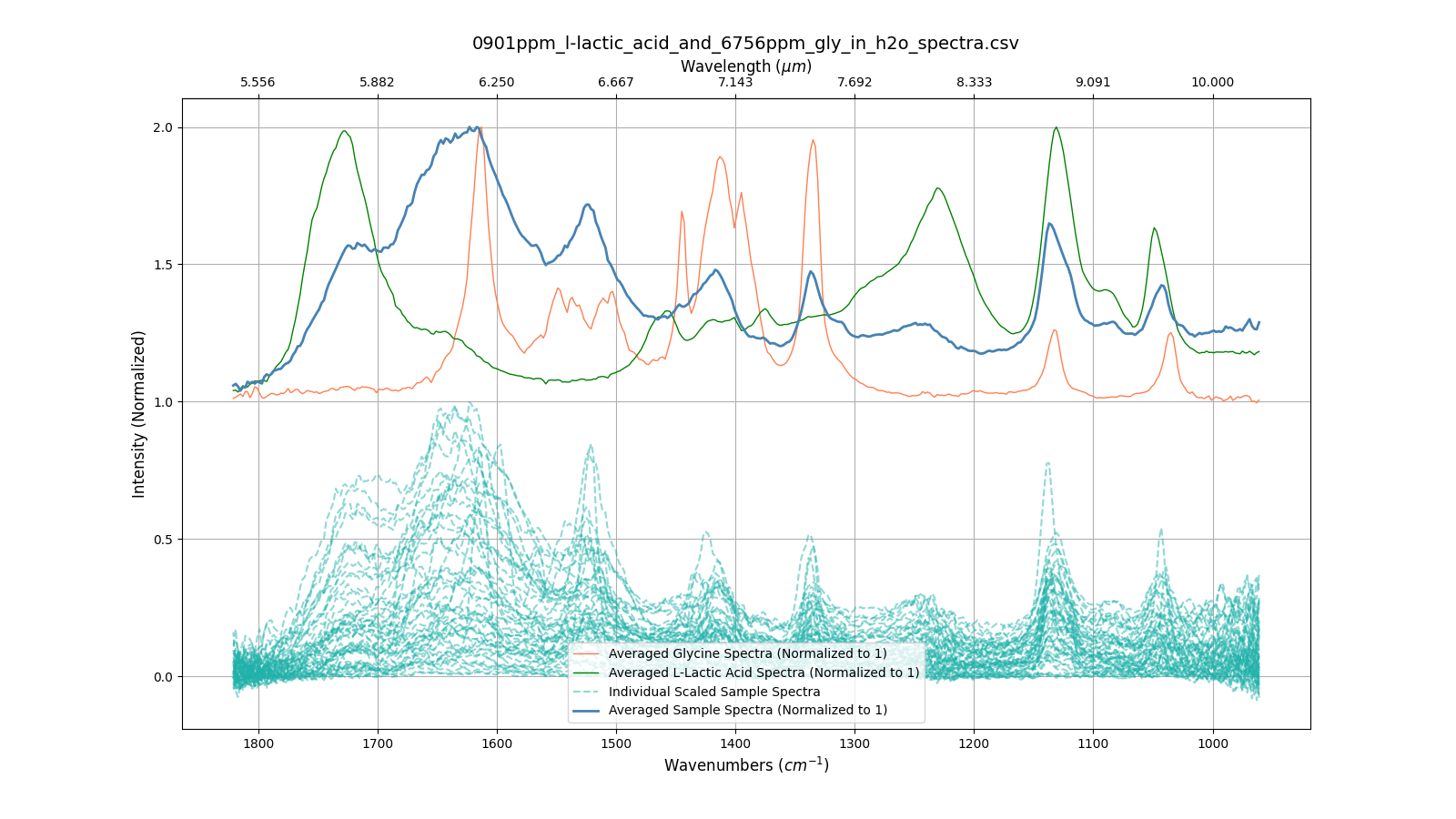}
    \captionof{figure}{901 ppm Lactic Acid / 6756 ppm L-glycine/Water Ice Mixture}
    \label{fig:Figure_11}
\end{center}

In Figure~\ref{fig:Figure_12} we show the other end of the mixture ratios with 8107 ppm of lactic acid and only 751 ppm of L-glycine. The lactic acid spectrum visible dominates the composite spectrum of the mixture as expected. Noticeably the water H-O-H bending moment peak is significantly reduced over the predominantly L-glycine mixture.

\begin{center}
    \includegraphics[width=0.48\textwidth]{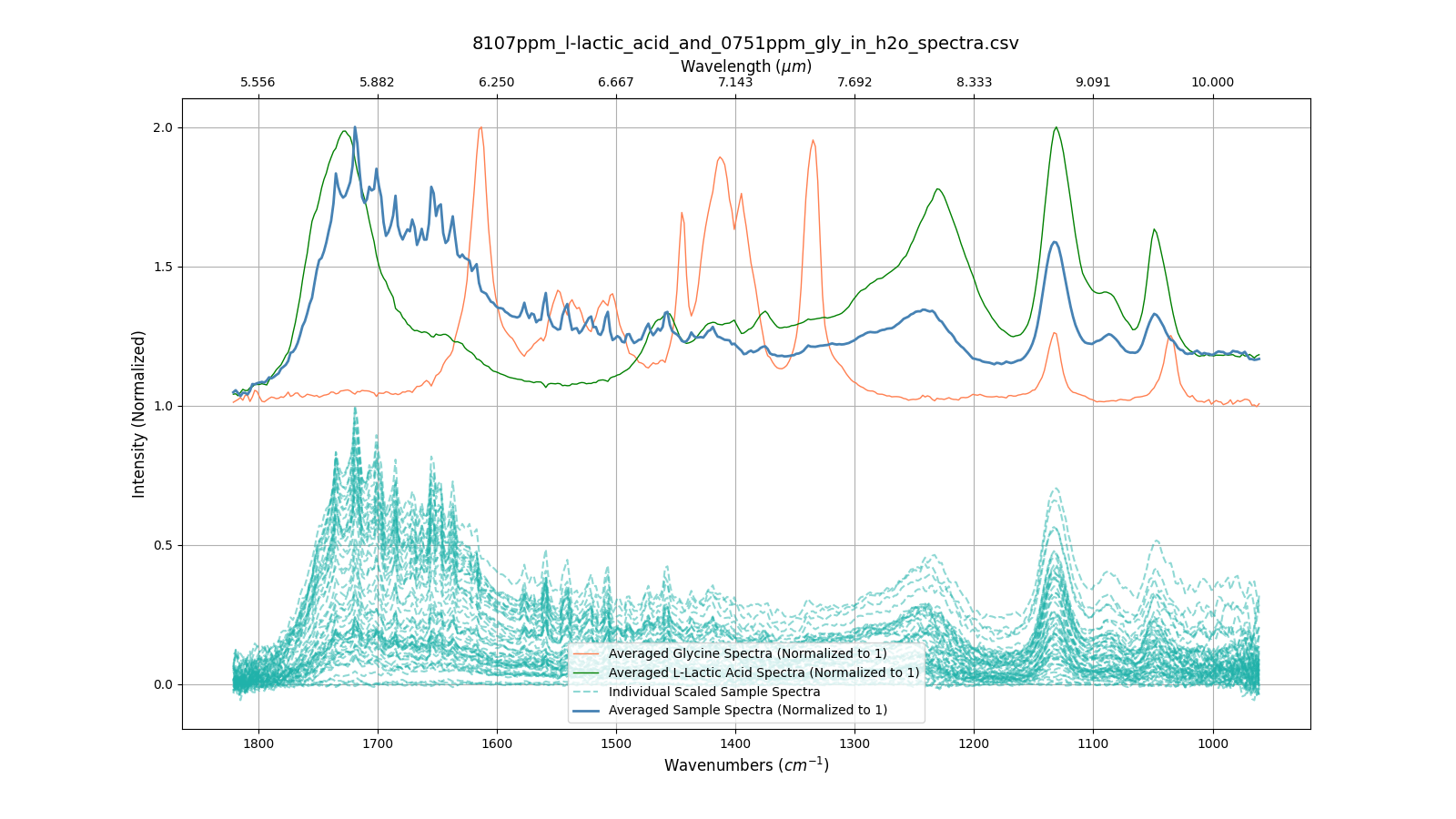}
    \captionof{figure}{8107 ppm Lactic Acid / 751 ppm L-glycine/Water Ice Mixture}
    \label{fig:Figure_12}
\end{center}

\subsubsection{Quantitative Ratio Determination}
We present the results of the analysis of the collected mixture data, applying the methods as described in \textbf{section 2.3.2}, to determine the calculated ratio of L-glycine to lactic acid from the spectral data. In Figure~\ref{fig:Figure_13} we show a comparison of the calculated ratios of the real data against the known prepared ratios in a synthetic model to establish the accuracy of the quantitative method. For this case with a large L-glycine to lactic acid ratio we see a broadening factor of B.F. = 1.03, this is consistent with a near perfect match (B.F. = 1.0) against an analytic model using the known mixture ratios.  

\begin{center}
    \includegraphics[width=0.48\textwidth]{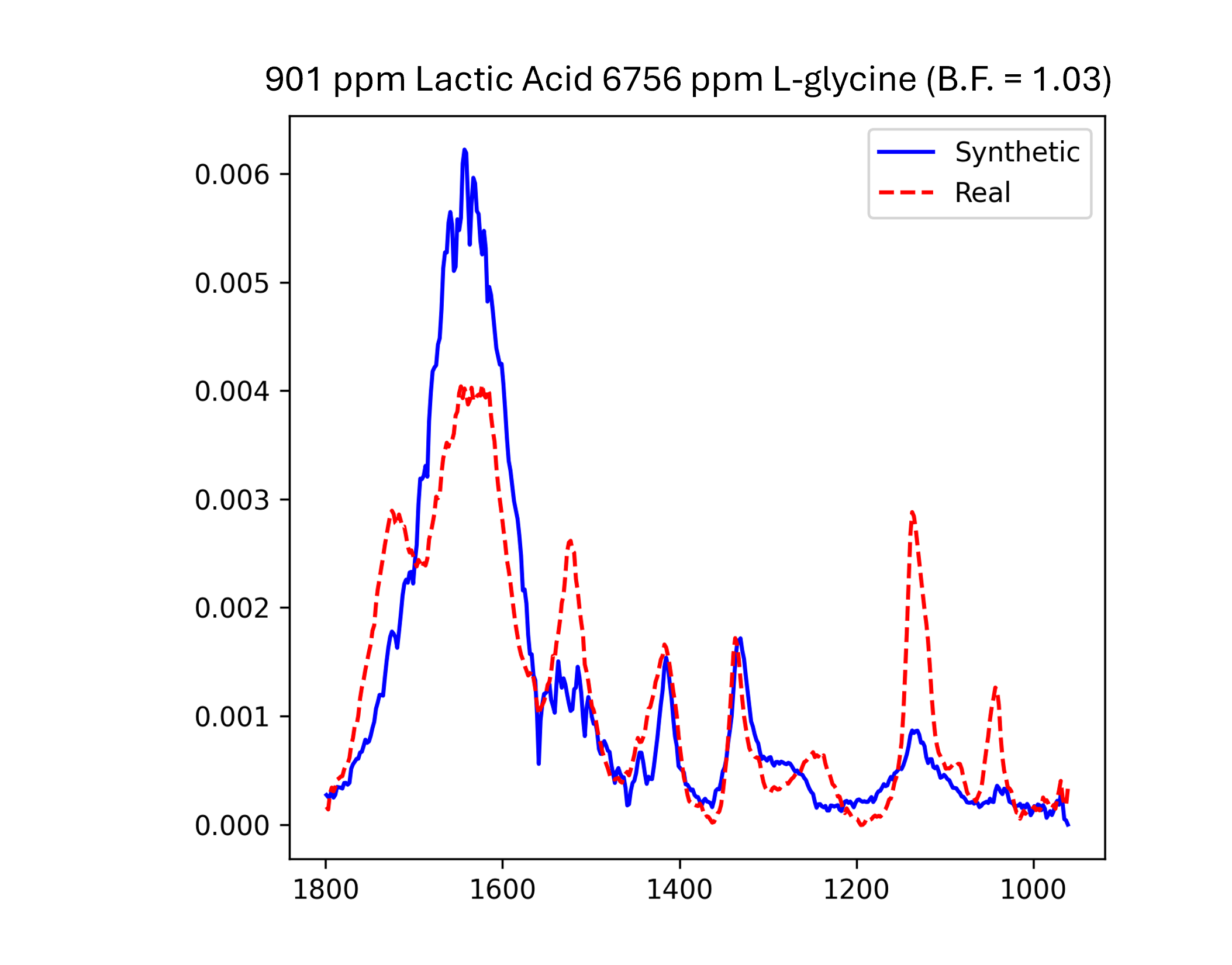}
    \captionof{figure}{901ppm Lactic Acid, 6756ppm L-glycine Real Data vs Synthetic Model}
    \label{fig:Figure_13}
\end{center}

In Figure~\ref{fig:Figure_14} we report the other mixture extreme, where the lactic acid dominates and shows a broadening factor of B.F. = 0.93. This indicates a good agreement between the mixture model, based solely on the concentration weighting, and the actual data.  Visually we can see the strong lactic acid peak shifted to the left, away from the H-O-H bending moment wavenumber and L-glycine peak.

\begin{center}
    \includegraphics[width=0.48\textwidth]{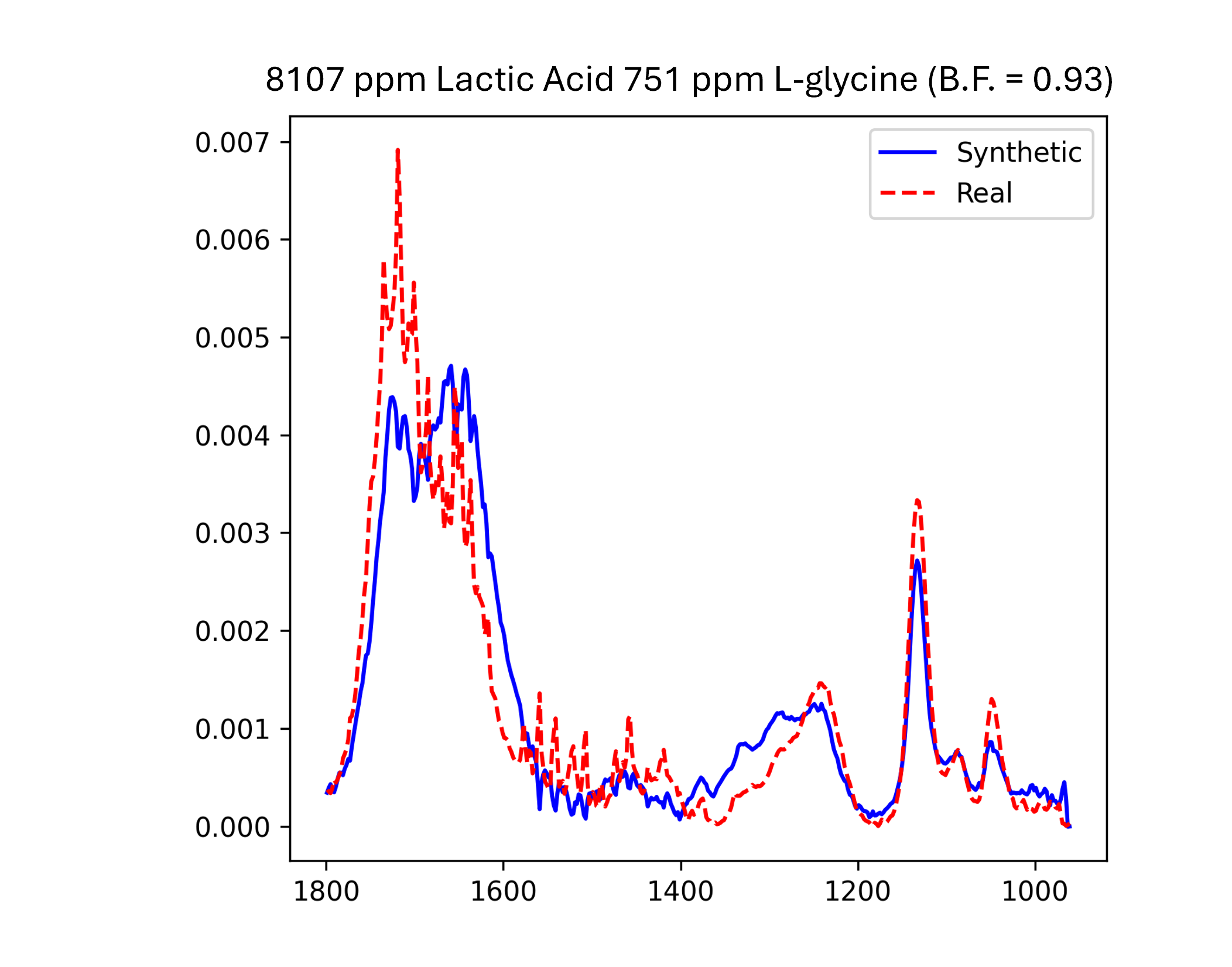}
    \captionof{figure}{8107ppm Lactic Acid, 751ppm L-glycine Real Data vs Synthetic Model}
    \label{fig:Figure_14}
\end{center}

Table~\ref{tab:mixture_analysis} shows the complete mixture results - we note that the broadening factor is near 1 (close agreement with synthetic model) for all of the samples except for the 4478 ppm lactic acid and 3732 ppm L-glycine mixture.  The B.F. value of 0.52 implies that the data exhibits a less broadened set of peaks, the graph has correctly aligned peaks and amplitudes, those peaks are just a sharper than the model suggest they should be.  To preclude an error in the process or methods we re-ran this particular data point two more times with identical results. This happens to be near a 1:1 ratio of the two organics, and could represent a specific co-crystalline structure or eutectic behavior at that specific concentration ratio. Such binary assemblies are well-documented in solid-state chemistry, where the zwitterionic properties and versatile hydrogen-bonding networks of amino acids make them highly effective co-formers in establishing ordered, multi-component crystal architectures (\cite{Nugrahani2021}.  Further study is needed to fully understand this outlier.

\begin{center}
\begin{table}[H]
\caption{Final Mixture Analysis Summary}
\label{tab:mixture_analysis}
\begin{tabular}{rccc}
\toprule
\textbf{Lactic Acid(ppm)} & \textbf{L-glycine(ppm)} & \textbf{R2} & \textbf{B.F.} \\
\midrule
901  & 6756 & 0.5817 & 1.03  \\
2252 & 5630 & 0.9471 & 0.86  \\
3603 & 4504 & 0.8428 & 0.93  \\
4478 & 3732 & 0.9388 & 0.52  \\
5405 & 3003 & 0.8942 & 0.90  \\
6756 & 1877 & 0.6860 & 1.07  \\
8107 & 751  & 0.8116 & 0.93  \\
\bottomrule
\end{tabular}
\end{table}
\end{center}

\subsubsection{Spatial Heterogeneity in Mixture Maps}
Using the O-PTIR systems integrated microscope we were able to visually analyze the ice surface and select the area for spectroscopic investigation. Following the collection we could view the hyperspectral maps to analyze the heterogeneous spatial distribution of each organic component within the frozen mixture, highlighting the value of sub-micron resolution. Figure~\ref{fig:Figure_15} shows an image of an organic ice surface captured from the O-PTIR high resolution microscope.

\begin{center}
    \includegraphics[width=0.40\textwidth]{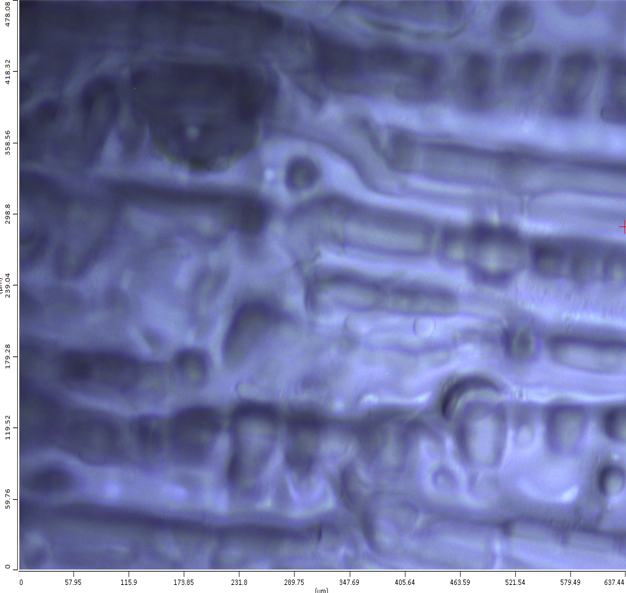}
    \captionof{figure}{Microscopic Image of Organic Ice Mixture}
    \label{fig:Figure_15}
\end{center}

Figure~\ref{fig:Figure_16} shows a hyperspectral surface map of an organic ice sample, the red areas indicate a higher signal for a specific wavenumber, the user can select this wavenumber filter to highlight specific peaks of interest associated with specific organic peaks.  The heatmap amplitudes align with the amplitudes of the peaks in the spectral data files.

\begin{center}
    \includegraphics[width=0.48\textwidth]{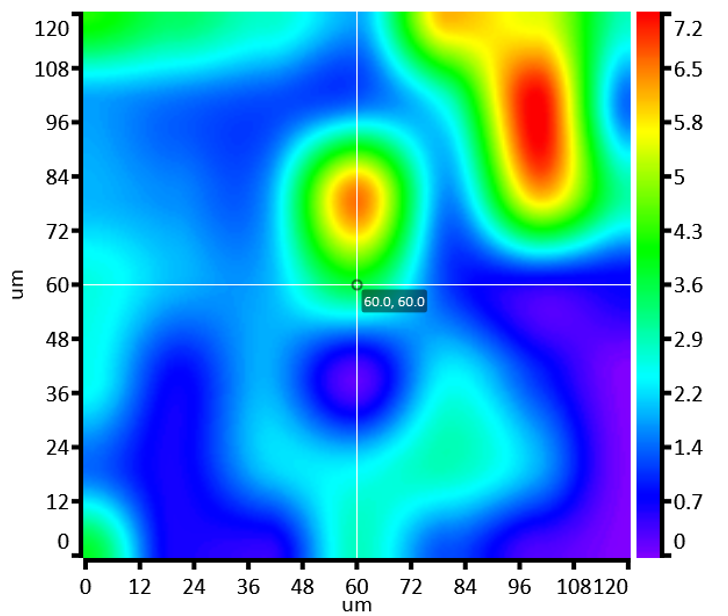}
    \captionof{figure}{Hyperspectral "Heatmap" of Organic Ice Mixture}
    \label{fig:Figure_16}
\end{center}

\section{Conclusions}
The O-PTIR system performed well to analyze both single and multiple component organic ice mixtures, with several amino acids (L-glycine and L-alanine) and an alpha-hydroxy acid, specifically lactic acid.  The instrument demonstrated detection limit, LOD, of $0.2 \mu M$ and a quantification limit, LOQ, of $3.83 \mu M$. Analyzing laboratory standard binary mixtures of organics (lactic acid and L-glycine), O-PTIR was able to distinguish the concentration of the organics in agreement with a simple analytic model which used a superposition of the organic spectrum weighted by their concentration percentage. This data collection was accomplished with very low laser powers and short collection times of ~45 min per data set, which included the time to auto-focus on the ice samples which are not flat.  O-PTIR appears to have utility and applicability to this type of data set and produces high quality, high spatial resolution data that would be invaluable as an instrument on a future in-situ collection mission and augment other more destructive analysis techniques that would otherwise lose the spatial context of the collections.

\section*{Declaration of Generative AI and AI-assisted Technologies in the
Manuscript Preparation Process}
During the preparation of this manuscript, the authors used generative AI tools, including Copilot (Microsoft) and Gemini (Google), to improve the readability and clarity of some sections. The authors performed a thorough review of all the manuscripts content and assume full responsibility for the scientific accuracy, originality, and integrity of the published article.

\section*{Declaration of Competing Interest}
The authors declare that they have no known competing financial
interests or personal relationships that could have appeared to influence
the work reported in this paper.

\section*{Data Availability Statement}

The O-PTIR spectroscopic data that supports the findings in this paper is available from the authors upon reasonable request

\section*{Acknowledgments}
This research was supported by NASA PICASSO Grant \#80NSSC22K1231 and the NASA Solar System Exploration Research Virtual Institute (SSERVI) under Cooperative Agreement No. NNH22ZDA020C (Center for Lunar Environment and Volatile Exploration Research - CLEVER), grant \#80NSSC23M0229.

\printcredits

\bibliographystyle{cas-model2-names}

\bibliography{references}

\end{document}